\documentstyle[epsf]{aipproc}

\def\ts{\textstyle}
\begin{document}

\catcode`@=11
\long\def\@makefntext#1{\parindent 0pt\hsize\columnwidth\parskip0pt\relax
\footnotesize\baselineskip12pt\def\strut{\vrule width0pt height0pt depth1.75pt\relax}%
\mbox{$\m@th^{\@thefnmark}$\hspace*{3pt}}#1}

\def\title#1{\gdef\@title{{\par\vskip-10pt\large\bf
\baselineskip20pt\centering\ignorespaces#1\vskip6pt}}%
\setcounter{part}{0}
\setcounter{table}{0}
\setcounter{figure}{0}
\setcounter{equation}{0}
\setcounter{section}{0}
\setcounter{subsection}{0}
\setcounter{subsubsection}{0}
\setcounter{paragraph}{0}
}

\def\author#1{\expandafter\def\expandafter\@authoraddress\expandafter
{\@authoraddress %
{\dimen0=-\prevdepth \advance\dimen0 by1.2\baselineskip
\nointerlineskip \centering
\vrule height\dimen0 width0pt\relax\ignorespaces\large\rm#1\par
}%
}%
}

\def\address#1{\expandafter\def\expandafter\@authoraddress\expandafter
{\@authoraddress{\nointerlineskip\vskip1pc
                 \footnotesize\it\centering\ignorespaces#1\par}}}

\catcode`@=12


\renewcommand{\thefootnote}{\fnsymbol{footnote}}

\font\fortssbx=cmssbx10 scaled \magstep2
\hbox to \hsize{
\includegraphics{uwlogo.ps}
\hskip.35in \raise.1in
\hbox{\fortssbx University of Wisconsin - Madison}
\hfill$\vcenter{\hbox{\bf MADPH-99-1103}
                \hbox{January 1999}}$ }

\title{\mbox{The Pattern of Neutrino Masses and How to Determine It}\footnote{Talk presented at the {\it 2nd International Workshop on Particle Physics and the Early Universe} (Cosmo~98), Asilomar, Monterey, CA, Nov.~1998.}}
\author{V. Barger}
\address{Physics Department, University of Wisconsin, Madison, WI 53706}
\maketitle

\section{Introduction}
Our knowledge of the neutrino sector of the Standard Model has recently undergone a revolution. Deficits of the atmospheric muon neutrino flux and the solar electron neutrino flux compared to their predicted values can be understood in terms of neutrino oscillations\cite{nu98} and we can therefore infer that neutrinos have non-degenerate masses. Additional but somewhat less secure evidence for $\bar\nu_\mu \to \bar\nu_e$ and $\nu_\mu\to\nu_e$ oscillations has been found in the LSND accelerator experiment\cite{nu98}. Because these experiments have widely different $L/E_\nu$ ranges ($\approx 10$ to $10^4$~km/GeV for atmospheric, $\approx 10^{11}$ for solar, and $\approx 1$ for LSND), the mass-squared differences required to explain the phenomena must be distinct. Given the observations, an important next step is to deduce the pattern of neutrino masses and mixings. Such studies depend on the number of neutrinos. The invisible width of the $Z$-boson measured in LEP experiments gives $N_\nu = 2.993\pm 0.011$, consistent with the usual $\nu_e, \nu_\mu$ and $\nu_\tau$ ``active" neutrinos. But there may also be right-handed ``sterile" neutrinos with no weak interactions. Only the observation of oscillations of the active neutrinos to sterile neutrinos can test for their existence. In the following, we first discuss the atmospheric and solar neutrino data in a 3-neutrino framework and then later generalize our considerations to include the LSND data with oscillations of four neutrinos.
The proceedings of the Neutrino~98 conference\cite{nu98}, a current review by the organizer of COSMO\,98\cite{caldwell}, and recent phenomenological analyses\cite{bahcall,bpww,bilenky,gibbons,fogli,hata,concha} may be consulted for references to the vast primary literature.

When neutrino flavor eigenstates $\nu_f$ are not the same as the mass eigenstates $\nu_i$, e.g., for two neutrinos, 
\begin{equation}
\nu_f = \cos\theta\nu_1 + \sin\theta\nu_2\,,\qquad
\nu_{f'} = -\sin\theta\nu_1 + \cos\theta\nu_2\,,
\end{equation}
then neutrinos oscillate. The vacuum oscillation probabilities are
\begin{eqnarray}
P(\nu_f\to\nu_{f'}, L) &=& A\sin^2\left(\delta m^2 L
\over 4E\right)\quad\qquad\mbox{``appearance''}\,,\\
P(\nu_f\to\nu_f, L) &=& 1 - A \sin^2 \left(\delta m^2 L
\over 4E\right)\qquad\mbox{``survival''}\,,
\end{eqnarray}
where $A=\sin^22\theta$, $\delta m^2 = m_2^2 - m_1^2$; $L$ is the path length and $E$ is the neutrino energy. The neutrino anomalies can be explained by effective two-neutrino oscillations; vacuum oscillations can account for the solar\cite{vlw}, atmospheric\cite{oldatmos} and LSND anomalies; matter-enhanced oscillations\cite{MSW} are an alternative to explain the solar neutrino deficit.

\section{Neutrino Anomalies and their Oscillation Interpretations}

\underline{\it Atmospheric} \ 
Cosmic ray interactions with the atmosphere produce $\pi$-mesons and the decays $\pi\to\mu\nu$ and $\mu\to\nu_e e\nu_\mu$ give $\nu_\mu$ and $\nu_e$ fluxes in the approximate ratio $(\nu_\mu+\bar\nu_\mu)/(\nu_e+\bar\nu_e)\sim2$ for $E_\nu\sim 1$~GeV. Measurements of $R=(N_\mu/N_e)_{\rm data} / (N_\mu/N_e)_{\rm MC}$ for $E_\nu\sim 1$~GeV find values of $R\sim 0.6$\cite{nu98}. In the water Cherenkov experiments the single rings from muons are fairly clean and sharp, while those from electrons are fuzzy due to electromagnetic showers. The separate distributions of $\mu$-like and $e$-like events versus the zenith angle establish that the anomalous $R$-ratio is due to a deficit of upward $\mu$-like events. As suggested long ago\cite{oldatmos}, the data are well described by $\nu_\mu\to\nu_\tau$ or $\nu_\mu\to \nu_s$ oscillations with $\delta m_{\rm ATM}^2 \approx 3\times10^{-3}\rm\,eV^2$ and $A_{\rm ATM}\approx1$.  For sub-GeV neutrino energies, $L/E$ is large at $\cos\theta<0$ and the oscillations average, $P(\nu_\mu\to\nu_\mu)\approx 0.5$. At multi-GeV energies, $L/E$ is large at $\cos\theta=-1$ and $P(\nu_\mu\to\nu_\mu)\approx 0.5$; also $L/E$ is small at $\cos\theta=+1$ and $P(\nu_\mu\to\nu_\mu)\approx 1$.

\underline{\it Solar} \
Three types of solar $\nu_e$ experiments, (i)~$\nu_e$ capture in Cl [Homestake], (ii)~$\nu_e e\to\nu_e e$ [Kamiokande and SuperKamiokande], (iii)~$\nu_e$ capture in Ga, measure rates below standard model expectations. The different experiments are sensitive to different ranges of solar $E_\nu$. Three regions of oscillation parameter space are found to accommodate all these observations\cite{bahcall,fogli,hata}:

\begin{center}
\begin{tabular}{lcc}
& $\delta m^2_{\rm SOL}\rm\ (eV)^2$& $A_{\rm SOL}$\\
Small Angle Matter (SAM)& $\sim 10^{-5}$& $\sim10^{-2}$\\
Large Angle Matter (LAM)& $\sim 10^{-5}$& $\sim 0.6$\ \ \ \\
Vacuum Long Wavelength (VLM)& $\sim 10^{-1\rlap{\scriptsize0}}$& $\sim1$\ \ \  \ \
\end{tabular}
\end{center}

\begin{figure}
\centering\leavevmode
\epsfxsize=2.75in\epsffile{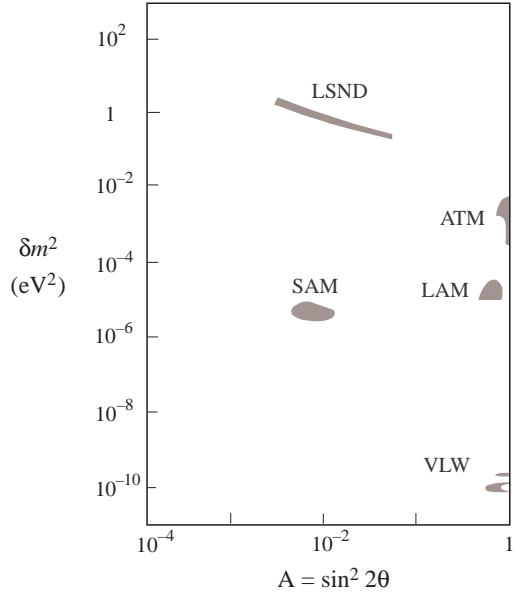}

\caption[]{Regions of oscillation parameters $\delta m^2, \sin^22\theta$ that can explain the atmospheric, solar, and LSND anomalies. \label{fig:regions}}
\end{figure}

\underline{\it LSND} \ 
The Los Alamos experiment studied $\bar\nu_\mu\to\bar\nu_e$ oscillations from $\bar\nu_\mu$ of $\mu^+$ decay at rest and $\nu_\mu\to\nu_e$ from $\nu_\mu$ of $\pi^+$ decay in flight. The results, including restrictions from BNL, KARMEN and Bugey experiments, suggest $\nu_\mu\to \nu_e$ oscillations with
\begin{equation}
0.3{\rm\ eV^2} < \delta m_{\rm LSND}^2 <  2.0{\rm\ eV^2}\,,\qquad
A_{\rm LSND} \approx 4\times10^{-2} \ \rm to \ 3\times10^{-3} \,.
\end{equation}
Figure \ref{fig:regions} illustrates the parameter regions for the solar, atmospheric and LSND oscillation interpretations. 
Since three distinct $\delta m^2$ are needed to explain the atmospheric, solar and LSND data, but there are only two independent $\delta m^2$ from $\nu_e, \nu_\mu, \nu_\tau$ neutrinos, there are two possible roads to follow: (i)~put the LSND anomaly aside until it is confirmed by the KARMEN and mini-BooNE experiments, or (ii)~explain all three anomalies by invoking oscillations to a sterile neutrino as well. We consider both routes in the following.

\section{New Inferred Limits on Neutrino Mass}

The fact that the neutron undergoes $\beta$-decay implies that the
electron-neutrino is a linear combination of one or more mass eigenstates with
mass below the kinematic $\beta$-decay end-point limit of $m_\beta=4.4$~eV.  The interpretation of the atmospheric and solar
neutrino anomalies in terms of oscillations tells us that the neutrino
mass splittings in a three-neutrino universe
are much smaller than $m_\beta$.  We conclude therefore that
all three neutrino mass eigenvalues satisfy $m_j \leq m_\beta$
and that the linear combinations of these mass states
which are $\nu_\mu$ and $\nu_\tau$ have effective masses $\leq m_\beta$ as well\cite{ourlimit}. These bounds represent a factor of $10^5$ to $10^6$ improvement over the current bounds $m_{\nu_\mu} < 170$~keV and $m_{\nu_\tau} < 18.2$~MeV. The largest mass eigenvalue is bounded below by $m_\beta \geq \sqrt{\delta m^2_{\rm atm}} \geq 0.002$~eV. Generalizing to include $\nu_\mu \rightarrow \nu_e$ oscillations in the LSND experiment with one sterile
and three active neutrinos, we obtain an upper bound
of 5.4~eV on all four neutrino masses and a lower bound on the largest mass eigenstate of $m_4 \geq \sqrt{\delta m^2_{\rm LSND}} \gtrsim 0.5$~eV.

\section{From Effective 2-Generation to 3-Generation Oscillations}

The neutrino MNS  mixing matrix\cite{kaoru-mns}, for either Dirac or Majorana neutrinos, can be parametrized by three angles $\theta_i$ and a CP-violating phase as
\begin{equation}
\left( \begin{array}{c} \nu_e \\ \nu_\mu \\ \nu_\tau \end{array} \right)
= U \left( \begin{array}{c} \nu_1 \\ \nu_2 \\ \nu_3 \end{array} \right)
= \left( \begin{array}{ccc}
  c_1 c_3                           & c_1 s_3
& s_1 e^{-i\delta} \\
- c_2 s_3 - s_1 s_2 c_3 e^{i\delta} &   c_2 c_3 - s_1 s_2 s_3 e^{i\delta}
& c_1 s_2 \\
  s_2 s_3 - s_1 c_2 c_3 e^{i\delta} & - s_2 c_3 - s_1 c_2 s_3 e^{i\delta}
& c_1 c_2 \\
\end{array} \right)
\left( \begin{array}{c}
\nu_1 \\ \nu_2 \\ \nu_3
\end{array} \right) \,,
\label{U}
\end{equation}
where $c_j \equiv \cos\theta_j$, $s_j \equiv \sin\theta_j$, and $\nu_1, \nu_2, \nu_3$ are the mass eigenstates. The vacuum oscillation probabilities of interest are
\begin{eqnarray}
\noalign{\hbox{Atmospheric:}}
P(\nu_\mu\rightarrow\nu_\mu) &=&
1 - (c_1^4 \sin^22\theta_2 + s_2^2 \sin^22\theta_1) \sin^2\Delta_{atm} \\
 P(\nu_e\rightarrow\nu_e) &=& 1 -
\sin^22\theta_1 \sin^2\Delta_{atm} \\
P(\nu_e\leftrightarrow\nu_\mu) &=&
s_2^2 \sin^22\theta_1 \sin^2\Delta_{atm} \\
P(\nu_e\leftrightarrow\nu_\tau) &=&
c_2^2 \sin^2 2\theta_1 \sin^2\Delta_{atm} \\
P(\nu_\mu\leftrightarrow\nu_\tau) &=&
c_1^4 \sin^2 2\theta_2 \sin^2\Delta_{atm} \\
\noalign{\hbox{Solar}}
P(\nu_e \rightarrow \nu_e) &=& 1 - \textstyle{1\over2}\sin^22\theta_1
- c_1^4\sin^22\theta_3 \sin^2\Delta_{sun} \hspace*{1.5in}
\end{eqnarray}
When $\theta_1=0$, the atmospheric and solar oscillations decouple and $\nu_e$ does not oscillate in atmospheric and long-baseline experiments. Best fits\cite{ourlimit} to the SuperK data are obtained with $\theta_1 = 0$, $\theta_2=\pi/4$ with $2\sigma$ bounds of $\theta_1 < 17^\circ$ and $|\theta_2 - 45^\circ| < 13^\circ$. The angle $\theta_3$ is determined  by the solar data (for the just-so\cite{vlw} or the MSW solutions\cite{MSW}, whichever is chosen by experiment). Thus we already have achieved the partial reconstruction of the neutrino MNS matrix! 

\section{Bi-Maximal Mixing Model}

If the atmospheric and solar data are both described by maximal mixing, then there exists a unique mixing matrix for three flavors\cite{bimax},
\begin{equation}
U = \left( \begin{array}{ccc}
 {1\over\sqrt2} & -{1\over\sqrt2} & \phantom{-}0 \\
 {1\over2}      &   \phantom{-}{1\over2} & -{1\over\sqrt2} \\
 {1\over2}      &   \phantom{-}{1\over2} & \phantom{-}{1\over\sqrt2} \\
\end{array} \right)
\end{equation}
The off-diagonal oscillation probabilites in this model are
\begin{eqnarray}
P(\nu_\mu\to\nu_\tau) &=& \sin^2\Delta_{\rm ATM} -\ts {1\over4}\sin^2\Delta_{\rm SUN}\\
P(\nu_e\to\nu_\mu) &=& \ts{1\over2}\sin^2\Delta_{\rm SUN}\\
P(\nu_e\to\nu_\tau) &=& \ts{1\over2}\sin^2\Delta_{\rm SUN}
\end{eqnarray}
where $\Delta \equiv 1.27\delta m^2 L/E$. A variety of models with nearly bi-maximal mixing have also been considered\cite{bimax}.

\section{Unified Models}

In unified models based on SO(10)\cite{babu}, SU(5)\cite{hagiwara}, flipped SU(5)\cite{ellis}, or anomalous U(1)\cite{ramond}, large $\nu_\mu\to\nu_\tau$ oscillations are accommodated but the small-angle MSW solar solution is required. This prediction is a clear distinction from the bi-maximal mixing model.

\section{Distinguishing Solar $\nu$-oscillation Scenarios}

The solar $\nu$ oscillation solutions will eventually be distinguished by use of all the following measurements: (i)~time-averaged total flux, (ii)~day-night dependence (earth-matter effects), (iii)~recoil electron energy spectra in $\nu e \to \nu e$ events), (iv)~seasonal variation, and (v)~the neutral-current to charged-current event ratio (SNO experiment).
The non-observation of a day-night effect has already ruled out substantial regions of the $\delta m^2, \sin2\theta$ parameter space\cite{ourlimit}. The electron energy distribution from recent SuperKamiokande data (708~days) now favor the vacuum long-wavelength interpretation and nearly exclude the MSW solutions\cite{smy} unless an enhanced hep flux component is involved\cite{bahcall-krastev}. 
 The day-night dependence due to the Earth is possible only for MSW solutions and has already ruled out substantial regions of otherwise allowed $\delta m^2, \sin^22\theta$ values.  A seasonal variation beyond the $1/r^2$ dependence of the flux is another characteristic of only long-wavelength vacuum oscillations and the 708-day SuperK data seem to show a seasonal effect above the eccentricity correction. The neutral current measurements of the SNO experiment will distinguish active from sterile oscillations.

\section{Four-Neutrino Options}

\begin{figure}[t]
\centering\leavevmode
\epsfxsize=1.7in\epsffile{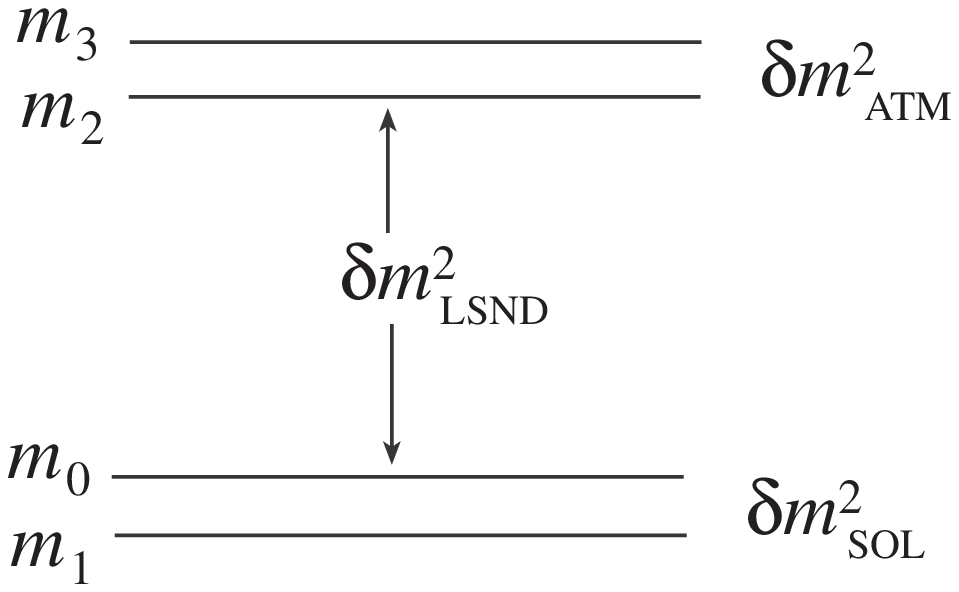}%
\hspace{.35in}\epsfxsize=2.7in\epsffile{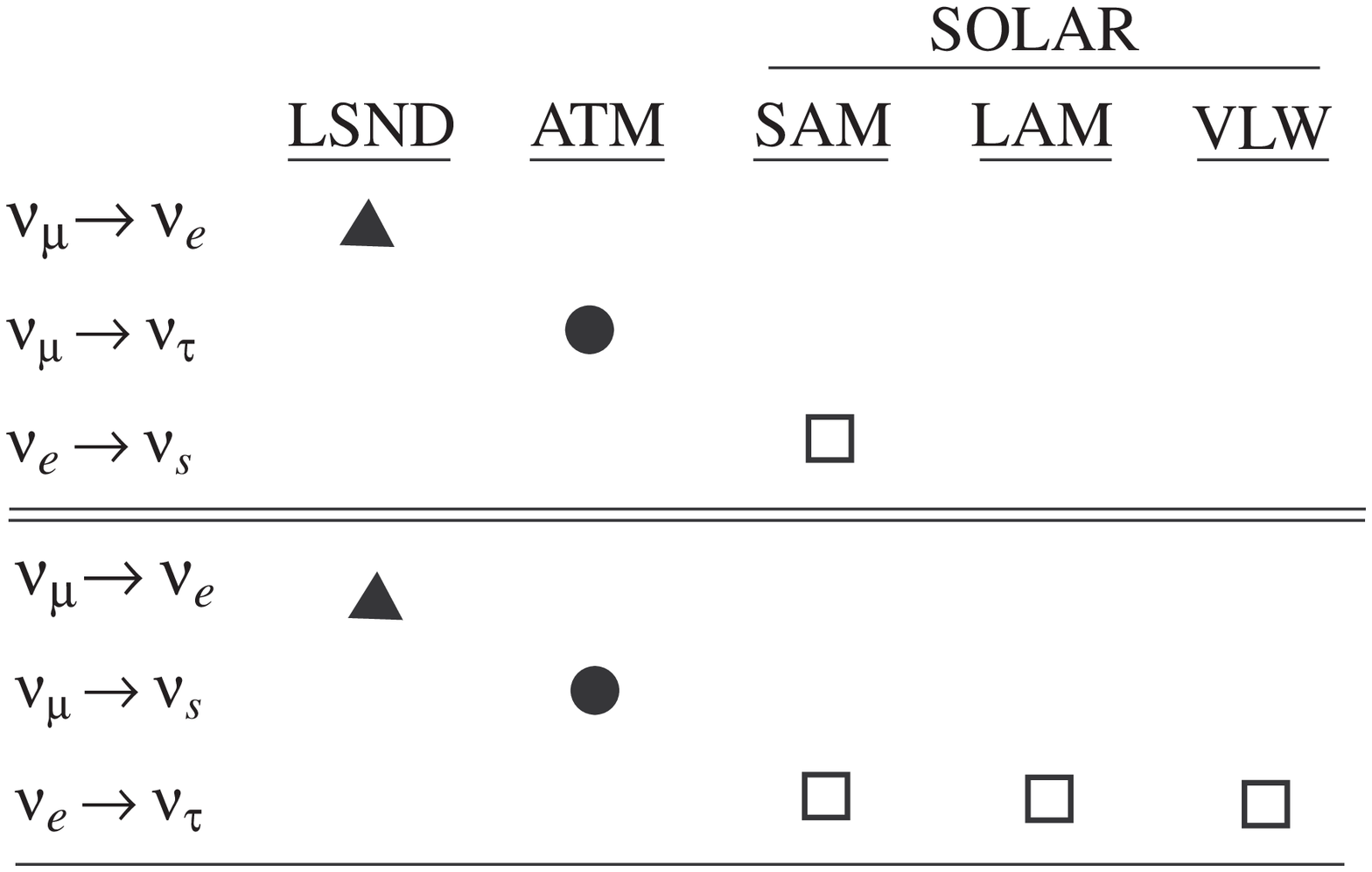}

\medskip
\caption[]{Neutrino mass spectrum showing which mass splittings are responsible for the LSND, atmospheric, and solar oscillations, and four-neutrino oscillation possibilities. \label{fig:spectrum}}
\end{figure}

With four neutrinos it is possible to account for the LSND data as well as the atmospheric and solar data. The preferred mass spectrum is two nearly degenerate mass pairs separated by the LSND scale\cite{bpww,bilenky}; see Fig.~\ref{fig:spectrum}. The figure also shows the options for oscillation solutions to all data. The alternative of a $1+3$ mass hierarchy with one heavier mass scale separated from three lighter, nearly degenerate states is disfavored when the null results of reactor and accelerator disappearance experiments are taken into account.

\noindent
Neutrino mass matrices have been proposed\cite{1096} that can account for the observations.

\section{Long-Baseline Experiments}

Long-baseline experiments with $L/E\approx 10$--$10^2$~km/GeV will measure $P(\nu_\mu\leftrightarrow\nu_\tau) \simeq \sin^2\Delta_{\rm ATM}$ and test the atmospheric oscillation result. In addition, the existence of $\nu_\mu\to\nu_e$ and $\nu_e\to\nu_\tau$ oscillations may be tested.
%
%
For this purpose intense neutrino beams are required. The MINOS experiment (Fermilab to Soudan) could confirm the SuperK $\nu_\mu\to\nu_\tau$ parameter region with $4\sigma$ sensitivty, provided that $\delta m_{\rm ATM}^2 > 2\times10^{-3}\rm\,eV^2$. The K2K experiment is sensitive to $\delta m^2 \gtrsim 3\times 10^{-3}\rm\,eV^2$.

In the future, a special purpose muon storage ring could provide high intensity neutrino beams with well-determined fluxes that could be directed towards any detector on the earth\cite{geer,bww}. It could be possible to store $\sim10^{21}$ $\mu^+$ or $\mu^-$ per year and obtain $\sim10^{20}$ neutrinos from the muon decays. Oscillations give ``wrong sign'' leptons from those produced by the beam. For example, $\mu^-$ decays give $\bar\nu_e$ and $\nu_\mu$ fluxes so detection of $\mu^+, e^-, \tau^\pm$ leptons tests for $\bar\nu_e\to\bar\nu_\mu(\bar\nu_\tau)$ and $\nu_\mu\to\nu_e(\nu_\tau)$ oscillations. Taus can be detected via their $\tau\to\mu$ decays and the $\tau$-charges so determined to distinguish $\nu_\mu\to\nu_\tau$ and $\bar\nu_e\to\bar\nu_\tau$ oscillations.


\begin{references}
\frenchspacing



\bibitem{nu98} Proceedings of Neutrino~98, Takayama, Japan [to be published], and references therein, 
http://www-sk.icrr.u-tokyo.ac.jp/nu98/scan/index.html

\bibitem{caldwell} D.O. Caldwell, IJMP {\bf A13}, 4409 (1998).

\bibitem{bahcall} J. Bahcall, P. Krastev, and A. Smirnov, Phys. Rev. {\bf D58}, 096016 (1998).

\bibitem{bpww} V. Barger, S. Pakvasa, T.J. Weiler, and K. Whisnant, Phys. 
Rev. {\bf D58}, 093016 (1998).

\bibitem{bilenky} S.M. Bilenky, C. Guinti, and W. Grimus, Eur. Phys. J. {\bf C1}, 247 (1998); C.~Guinti, C.W.~Kim, and M.~Monteno, Nucl. Phys. {\bf B521}, 3 (1998).

\bibitem{gibbons} D.O.~Caldwell and R.~Mohapatra, Phys. Rev. {\bf D48}, 3259 (1993); S.C.~Gibbons, R.N.~Mohapatra, S.~Nandi, and A.~Raychoudhuri, Phys. Lett. {\bf B430}, 296 (1998); S.~Mohanty, D.P.~Roy, and U.~Sarkar, hep-ph/9810309.

\bibitem{fogli} G.L. Fogli, E. Lisi, A. Marrone, and G. Scioscia, Phys. Rev. {\bf D59}, 033001 (1999).

\bibitem{hata} N. Hata and P. Langacker, Phys. Rev. {\bf D56}, 6107 (1997).

\bibitem{concha} M.C.~Gonzalez-Garcia, H.~Nunokawa, O.~Peres, T.~Stanev, and J.W.F.~Valle, Phys. Rev. {\bf D58}, 033004 (1998).

\bibitem{vlw}
V. Barger, R.J.N. Phillips, and K. Whisnant, Phys. Rev. {\bf D24},
538 (1981);
S.L. Glashow and L.M. Krauss, Phys. Lett. {\bf B190}, 199 (1987).

\bibitem{oldatmos}
J.G. Learned, S. Pakvasa, and T.J. Weiler, Phys. Lett. {\bf B207}, 79 (1988);
V. Barger and K. Whisnant, Phys. Lett. {\bf B209}, 365 (1988);
K. Hidaka, M. Honda, and S. Midorikawa, Phys. Rev. Lett. {\bf 61}, 1537
(1988).

\bibitem{MSW}
L. Wolfenstein, Phys. Rev. {\bf D 17}, 2369 (1978);
S.P. Mikheyev and A. Smirnov, Yad. Fiz. {\bf 42}, 1441 (1985);
Nuovo Cim. {\bf 9C}, 17 (1986);
S.P. Rosen and J.M. Gelb, Phys. Rev. {\bf D 34}, 969 (1986);
S.J. Parke, Phys. Rev. Lett. {\bf 57}, 1275 (1986);
W.C. Haxton, Phys. Rev. Lett. {\bf 57}, 1271 (1986);


\bibitem{ourlimit}
V. Barger, T.J. Weiler and K. Whisnant, Phys. Lett. {\bf B442}, 255 (1998).

\bibitem{kaoru-mns}
Z. Maki, M. Nakagawa and S. Sakata, Prog. Theor. Phys. {\bf 28}, 870 (1962).


\bibitem{bimax}
V. Barger, S. Pakvasa, T.J. Weiler, and K. Whisnant,
Phys. Lett. {\bf B437} 107 (1998);
A.J. Baltz, A.S. Goldhaber, and M. Goldhaber, Phys. Rev. Lett. {\bf 81}, 5730 (1998);
M. Jezabek and Y. Sumino, Phys. Lett. {\bf B440}, 327 (1998);
Y. Nomura and T. Yanagida, Phys. Rev. {\bf D59}, 017303 (1999);
G. Altarelli and F. Feruglio, Phys. Lett. {\bf B439}, 112 (1998);
JHEP {\bf 9811}, 021 (1998);
H. Georgi and S. Glashow, hep-ph/9808293;
S. Davidson and S.F. King, Phys. Lett. {\bf B445}, 191 (1998).
R.N. Mohapatra and S. Nussinov, Phys. Lett. {\bf B441}, 299 (1998); hep-ph/9809415;
S.K. Kang and C.S. Kim, hep-ph/9811379;
H. Fritzsch and Z. Xing, Phys. Lett. {\bf B372}, 265 (1996);
hep-ph/9807234; Phys. Lett. {\bf B440}, 313 (1998);
E. Torrente-Lujan, Phys. Lett. {\bf B389}, 557 (1996);
M. Fukugita, M. Tanimoto, and T. Yanagida, Phys. Rev. {\bf D57}, 4429 (1998);
M. Tanimoto, Phys. Rev. {\bf D59}, 017304 (1999);
Y.-L.~Wu, Academia Sinica report AS-ITP-99-04 [hep-ph/9901230].


\bibitem{babu} J.K. Elwood, N. Irges, and P. Ramond, Phys. Rev. Lett. {\bf 81}, 5064 (1998); C.H.~Albright, K.S.~Babu, and S.M.~Barr, Phys. Rev. Lett. {\bf 81}, 1167 (1998); K.S. Babu, J.C.~Pati, and F.~Wilczek, report OSU-HEP-98-11 [hep-ph/9812538]. 

\bibitem{hagiwara} K. Hagiwara and N. Okamura, report KEK-TH-605 [hep-ph/9811495].

\bibitem{ellis} J. Ellis, G.K. Leontaris, S. Lola, and D.V. Nanopouos, report ACT-7-98 [hep-ph/9808251].

\bibitem{ramond} J. Harvey, D.B. Reiss, and P. Ramond, Nucl. Phys. {\bf B199}, 223 (1982);
.


\bibitem{smy} M.B. Smy, talk presented at the DPF\,99 meeting, UCLA, 1999; G.~Sullivan, talk presented at the Aspen Winter Conference on Particle Physics, Aspen, CO, Jan.~1999.

\bibitem{bahcall-krastev} J.N. Bahcall and P. Krastev, Phys. Lett. {\bf B436}, 243 (1998).


\bibitem{1096} V. Barger, Y.-B.~Dai, K. Whisnant, and B.-L.~Young, University of Wisconsin-Madison report MADPH-99-1096 [hep-ph/9901388], and references therein.

\bibitem{geer} S. Geer, Phys. Rev. {\bf D57}, 6989 (1998).

\bibitem{bww} V. Barger, T.J. Weiler, and K. Whisnant, Phys. Lett. {\bf B427}, 97 (1998).

\end{references}
\end{document}